\documentclass[prb, onecolumn, showpacs, preprintnumbers, amsmath, amssymb, superscriptaddress, aps]{revtex4}
\usepackage{graphicx}
\usepackage{dcolumn}
\usepackage{color}
\usepackage{bm}
\def\he4{$^4$He}
\def\hee3{$^3$He}
\def\Am3{\AA$^{-3}$}
\def\beq{\begin{equation}}
\def\eeq{\end{equation}}
\begin{document}
%%%%%%%%%%%%%%%%%%%%%%%%%%%%%%%%%%%%%% AUTHORS %%%%%%%%%%%%%%%%%%%%%%%%%
\author{D. Aleinikava}

\author{E. Dedits}

\author{A.B. Kuklov}
\affiliation{Department of Engineering Science and Physics,
CSI, CUNY, Staten Island, NY 10314, USA}

\title{Glide and Superclimb of Dislocations in Solid $^4$He}

\begin{abstract}
Glide and climb of quantum dislocations under finite external stress, variation of chemical potential and bias (geometrical slanting) in Peierls potential are studied by Monte Carlo simulations of the effective string model. We treat on unified ground 
quantum effects at finite temperatures $T$. Climb at low $T$ is assisted by superflow along dislocation core -- {\it superclimb}. 
Above some critical stress  avalanche-type creation of kinks is found. It is characterized by hysteretic behavior at low $T$. 
At finite biases gliding dislocation remains rough even at lowest $T$ -- the behavior opposite to non-slanted dislocations.
In contrast to glide, superclimb is characterized by quantum smooth state at low temperatures even for finite bias. In some intermediate $T$-range giant values of the compressibility as well as non-Luttinger type behavior of the core superfluid are observed.

\end{abstract}

\maketitle

\section{Introduction}
Superfluid and mechanical properties of solid \he4 continue to be fascinating and, in many respects, largely enigmatic 
subjects for study. This is especially so for the proposed supersolid phase ~\cite{Andreev69} and the observed anomaly interpreted as non-classical rotational inertia (NCRI) in torsional oscillator (TO) experiments ~\cite{KC,KC2}. First-principle simulations ~\cite{noSFS} as well as a strong dependence of the NCRI on preparation and annealing protocols (see in Ref.~\cite{Balibar}) have definitely excluded supersolidity of ideal {\it hcp} solid \he4. Thus, the superfluid response of solid \he4 can only be due to superfluidity along extended structural defects forming a percolating network. Dislocations are one of the possible candidates for such network -- some of them do have superfluid cores as determined in {\it ab initio} simulations \cite{screw,stress,sclimb}.

Dislocations control mechanical properties of solids. One of the remarkable effects -- temperature induced shear modulus softening \cite{Friedel} has recently been observed in solid \he4 as well \cite{Beamish}. There is the whole plethora of very intriguing and largely unexplained features exhibited by the modulus at extremely small external stresses \cite{Beamish}. Very recently the opposite behavior -- low-$T$ softening has been reported \cite{Balibar3} for impurities free crystals.  

Superclimb of edge dislocations \cite{sclimb} is macroscopic quantum phenomenon where high-$T$ diffusion along dislocation core is replaced by superfluid transport. Convincing evidence for such transport has been presented in Refs.\cite{Ray}. There are three crucial aspects of the observation of mass transport through solid \he4 \cite{Ray}: i) Suppression above temperature $T_0 \approx 0.6$K which is significantly lower than that for the bulk $\lambda$ point; ii) Uniform transfer of time dependent pressure through the whole sample during the superflow events ; iii) Macroscopically large mass accumulation in the bulk during the superflow events. The feature i) clearly excludes any macroscopic superfluid channels scenario. The item ii) is inconsistent with classical plasticity driven by gradients of stress. The effect iii) finds its natural explanation in terms of superfluid mass transport along edge dislocation cores which adds (or removes) material to (from) half-plane forming edge dislocations \cite{sclimb}. 

In this paper, we analyze both aspects of dislocation thermodynamics -- their glide and superclimb. Specifically, we consider stress induced roughening of gliding dislocation in Peierls potential. Finite biases due to pinning of dislocation ends in different valleys of Peierls potential are also analyzed. We also consider temperature dependence of so called \textit{isochoric compressibility} $\chi(T)$ and specifically address the prediction of its suppression at low $T$ made in Ref.\cite{sclimb}. We propose utilizing two types of responses in experimental studies -- full and differential.

\section{String Model of Dislocation}
We model dislocation as a quantum string of length $L_x$ strongly pinned at its both ends and subjected to Peierls potential $U_P$, external force $F$ and slanting (bias).
Such string can perform two types of motion -- glide and superclimb. The model for glide has been discussed in detail in Ref.\cite{EPL}. The modification of this model
to describe superclimb has been introduced in Ref.\cite{sclimb}. Here we present unified description suitable for both cases, beginning with the superclimbing dislocation.
After trivial simplification, the resulting action can be utilized for glide as well.  In this paper we do not consider more involved model -- dislocation performing glide and superclimb.

Thermodynamical properties of dislocation are described by its partition function $Z=\int Dy(x,t) D\rho(x,t) D\phi(x,t) \exp(-H)$ calculated in imaginary time $0\leq t \leq \beta=1/T$ (units $\hbar=1$, $K_B=1$), where the functional integral is evaluated with respect to the dislocation displacement $y(x,t)$, density of the core $ \rho(x,t)$ and superfluid phase $\phi(x,t)$ conjugate to $\rho$, with $x$ being coordinate along the dislocation. The action 
\begin{eqnarray} 
H&=&H_{SF} + H_{dis} 
\label{H} \\
H_{SF}&=& \int_0^\beta dt\sum_{x}[ i(\rho +n_0)\nabla_t \phi  + \frac{\rho_0}{2} (\nabla_x\phi)^2  +\frac{1}{2\rho_0} (\rho - y)^2]
\label{HSF} \\
H_{dis} &=& \int_0^\beta \sum_{x}[ {1\over{2K_d}} \left((\nabla_t y)^2 + V_d^2(\nabla_x y)^2 \right) - u_P \cos\left(2\pi y\right)
\nonumber \\ 
&+&\frac{u_0}{2}y^2-F y(x,t)]
\label{Hdis}\\
y(0,t)&=&0; \, y(L_x,t)=y_{max}=0,1,2,..., \quad y(x,t+\beta)=y(x,t). %\quad B= 2\pi %4\sqrt{\pi}
\label{or}
\end{eqnarray}
is in continuous time $t$ and discrete space $x=0,1,2,...L_x$.
Here $H_{SF}$ describes core superfluidity \cite{sclimb}, with $n_0, \rho_0$ standing for average filling factor and bare superfluid stiffness, respectively; units are such that 
speed of sound in superfluid is unity. In this work we are considering the case of integer $n_0$. Non-integer $n_0$ lead to phase separation effects at low $T$ and require
additional analysis which will be presented elsewhere. 

The derivative $\nabla_x \phi$ is understood as a finite difference defined modulo $2\pi$: that is,
$\nabla_x \phi(x,t) =\cos( \phi(x+1,t) -\phi(x,t)) $. In what follows, $\cos$ will be treated in the Villain approximation 
$\cos( \theta(x,t)) \to \theta(x,t) + 2\pi m_x(x,t)$, 
with the ensemble summation over the integer variable $m_x(x,t)=0, \pm 1,\pm 2, ...$ performed. 

We discretize time into $N_t$ slices and replace 
$\nabla_t \phi(x,t) \to [\phi(x,t+1)-\phi(x,t) + 2\pi m_t(x,t)]\tau_0^{-1}$, with $m_t(x,t)=0,\pm 1,\pm 2, ...$, and $\tau_0$ standing for the time slice chosen as $\tau_0=1/(TN_t)$. The last term in Eq.(\ref{HSF}) describes the superclimb effect \cite{sclimb} -- building the edge of the dislocation so that its no energy-cost displacement $y\to y \pm 1, \pm 2, ...$ is possible by delivering matter $\rho \to \rho \pm 1,\pm 2, ...$, respectively, through superflow along the core. All the variables are periodic along the imaginary time  $\beta$. We consider dislocation attached to large superfluid reservoirs at its both ends. This implies periodic in space boundary conditions for supercurrents. 

The term (\ref{Hdis}) describes string motion, with $1/K_d, V_d$ standing for dislocation effective mass density and speed of sound (in units of speed of sound in superfluid), respectively, and $u_p$ denotes strength of Peierls potential, with Burgers vector $b$ taken as unity. 
The term $\sim u_0$ accounts for effects of dislocation pinning by \hee3 impurities (not discussed here) and various structural defects. 
The last two terms in Eq.(\ref{Hdis}) account for the effect of external stress $\sigma$ so that the linear force density $F\approx b\sigma$ (where we ignore spatial indices).  
The last line, Eq.(\ref{or}), describes strong pinning at $x=0$ and $x=L_x$ so that, in general, the string is biased by $y_{max}\neq 0$ to have finite number of kinks for gliding dislocation (and -- jogs for superclimbing) even at $T=0$.

We treat the Peierls term in the Villain approximation as well, that is, 
 $ - \cos\left(2\pi y(x,t)\right) \to \sim (y(x,t) + P(x,t))^2$ with the ensemble
summation over the discrete variable $P(x,t)=0, \pm 1, \pm 2,...$. This allows exact integrating out the variables $\rho, \,\, \phi, \,\, y$ analytically and expressing the partition function solely in terms of the integer variables $m_x(x,t),\, m_t(x,t),\, P(x,t)$. Integration of the phase generates the constraint
$\nabla_t m_t + \nabla_x m_x=0$, where now $\nabla_t,\, \nabla_x$ are understood as simple discrete differences with respect to the integer variables $m_x, m_t$ defined on the bonds of the space-time lattice. Such constraint constitutes so called J-current model \cite{Wallin} which was treated within the Worm Algorithm (WA) \cite{WA}. The variable $P$ is unconstrained. Monte Carlo simulations have been performed for fixed number of time slices $N_t$ (we used $N_t=40$ in this work) with updates typical for WA modified in the presence of independent integer variable $P$ defined on each site of the space-time lattice. Discrete time simulations carry systematic error decreasing with increasing $N_t$ (for fixed $T$). However, we found that thermal profiles of measured quantities are self-similar for different $N_t>20$ and, therefore, such error can be absorbed into the definition of unit of temperature. The value $N_t=40$ is a reasonable choice between acceptable systematic error and realistic simulation times. 

\subsection{Long-range interaction}

The parameter $K_d$ in Eq.(\ref{Hdis}) is not actually a constant. As discussed in Ref.\cite{EPL}, it contains a contribution due to long-range forces between kinks. At long distance $r$ between two kinks the interaction potential is $\sim 1/r$. Accordingly, $1/K_d$ has a log-divergent factor \cite{Hirth1,EPL}. Ignoring retardation effects (which do not lead to any significant differences), in Fourier
\begin{eqnarray}
K^{-1}_d(q) =  \frac{1}{K} C(q), \,\,\quad C(q)= 1+ U_C \ln\left(1+ \frac{1}{(b_0q)^2}\right),
\label{C} 
\end{eqnarray}
where $K\sim 1, U_C\sim 1$ are parameters; $b_0\sim 1$ stands for the short scale cutoff; $q$ is spatial Fourier vector along the dislocation. As discussed in Ref.\cite{EPL}, in the absence of the long-range forces, $U_C=0$, the parameter $K$ determines if the system is quantum rough (for $K>K_c$ with $K_c=2/\pi$) or quantum smooth ($K<K_c$). At $K=K_c$ it undergoes Berezinskii-Kosterlitz-Thouless (BKT) transition at $T=0$ driven by quantum fluctuations (for long strings). In the quantum rough state, Peierls potential is irrelevant and dislocation behaves as free Granato-L\"ucke string \cite{Granato}. One can estimate (in the standard units up to a numerical factor $\sim1$) $K=\hbar a /(m_0 V_db^2)$, where $m_0$ is atomic mass, $a$ stands for interatomic distance along the dislocation core; $b$ - Burgers vector; $V_d$ - speed of sound in standard units.  Evaluation of $K$ for solid \he4, shows that $K\sim 1$, that is, the situation is marginal (as compared to crystals of heavier elements or those with significantly higher $V_d$ where $K<<1$). However, as proven in Ref.\cite{EPL}, arbitrary small long-range interaction $U_C\neq 0$ always keeps unbiased dislocation in the smooth state at $T=0$. 
In contrast to glide, superclimb is not affected qualitatively by such long-range interaction \cite{sclimb}, and dislocation becomes quantum smooth at low $T$ for any value of $K$ even for $U_C=0$.   

\section{Glide}
In this section we will discuss gliding dislocation at finite external stress and bias.
The above model can be reduced to describe glide as considered in Ref.\cite{EPL} by setting $\rho_s \to \infty$ so that $H_{SF}$ can be simply set to zero.

We introduce two responses, one is differential 
\begin{eqnarray}
R_2= \frac{1}{\beta L_x} \int_0^\beta dt \sum_x \frac{d\langle y(x,t)\rangle}{dF} =\frac{1}{\beta L_x} \frac{d^2\ln Z(F)}{dF^2}
\label{R2}
\end{eqnarray}
and another one (we call it full)
\begin{eqnarray}
R_1= \frac{1}{\beta L_x} \int_0^\beta \sum_x \frac{\langle y(x,t)\rangle}{F} =\frac{1}{\beta L_x} \frac{d\ln Z(F)}{FdF}
\label{R1}
\end{eqnarray}
where the averages are taken over the full action (\ref{H}) with finite $F$ and bias $y_{max}$. Since at finite biases there is already finite $\langle y\rangle \sim y_{max}$, we modify Eq.(\ref{R1}) by requiring that $R_1$ includes average of two cases $\pm y_{max}$. This requirement reflects a generic situation that actual samples contain on average equal amount of oppositely slanted dislocations. In what follows we will see that, while at small stresses these two responses coincide, at finite stresses they behave quite differently, and, therefore, important information can be inferred about state of dislocations. Physically, two responses can be measured at slightly different frequencies so that total applied stress $\sigma=\sigma_0 + \sigma'$ consists of finite part $\sigma_0$ at one frequency and small addition $\sigma'$ -- at another. Then,
$R_2$ can be viewed as linear response on $\sigma'$ at finite $\sigma_0$. 

$R_{1,2}$ can be related to, respectively, full $G_1$ and differential $G_2$ shear moduli following the procedure described in detail in Ref.\cite{Granato}.
 If the dislocation network is viewed as a set of blocks of free segments of sizes $L_x,L_y,L_z$ along the corresponding orthogonal axes, a displacement $y$ (along $Y$-axis) under a  force $F=\sigma_{zy} b$  where $\sigma_{zy}$ stands for the $zy$-component of the stress tensor, results in a strain of the block $u_{zy}\approx b y/(L_yL_z)$ with $ y \propto \sigma_{zy}$. Specifically: 
\begin{eqnarray}
u_{zy}&=& \frac{bN_d}{\beta L_x} \int^\beta_0 dt \sum_x \langle y(x,t) \rangle,\,\quad N_d\equiv\frac{1}{L_yL_z},
\label{u_y}
\end{eqnarray}
where $N_d$ can be called dislocation density with respect to chosen coordinates.
We introduce dimensionless dislocation density $n_d\equiv b^2N_d$ in units of $b$. Clearly, our description is based on the assumption $n_d<<1$. It should also be noted that, in reality, dislocation density is not a scalar. We, however, will be working within the simplifying assumption $L_y\approx L_z \approx L_x$, so that $n_d\approx b^2/L_x^2$ can be treated as a scalar for all practical purposes. Eq.(\ref{u_y}) describes macroscopic plastic deformation. Adding elastic strain $u^{(el)}_{zy}=G^{-1}_e \sigma_{zy}$, where $G_e$ denotes bare elastic shear modulus, and relating full strain to stress, one finds 
\begin{eqnarray}
\frac{1}{G_1} = \frac{1}{G_e} + n_d R_1, \quad \frac{1}{G_2} = \frac{1}{G_e} + n_d R_2,
\label{G12}
\end{eqnarray}

At high $T$ Peierls potential becomes irrelevant, and the responses $R_{1,2}$ become equal to each other and to that of free string $R_0\approx (1/L_x)\sum_x 0.5 K V_d^{-2} x(L_x-x) \approx  KV_d^{-2} L_x^2/12$, where we ignored log-factor (\ref{C}). Thus, given $G_e KV^{-2}_d \approx 1$ and that $n_dL_x^2\approx 1$ for isotropic dislocation forest, one concludes that high-T values $G_{1\infty}=G_{2\infty}=G_{\infty} \approx G_e/(1+\gamma)$ where $\gamma \approx 0.1$ is a geometrical factor independent of dislocation density. Should, however, the dislocation forest has some asymmetry, say, $L^2_x >> L_yL_z$, the factor $\gamma \approx 0.1L_x^2/L_yLz$ can, in principle, take arbitrary large value. It is convenient to introduce $R_{1,2}$ relative to $R_0$ for given $L_x$. So, in what follows we will be discussing namely such values $\tilde{R}_{1,2}=R_{1,2}/R_0$, approaching unity at high $T$. Then, Eqs.(\ref{G12}) can be rewritten as
\begin{eqnarray}
G_{1,2} = \frac{G_e}{1 + \gamma \tilde{R}_{1,2}}
\label{G122}
\end{eqnarray}
\begin{figure}
\begin{center}
\includegraphics[%
  width=0.6\linewidth,
  keepaspectratio]{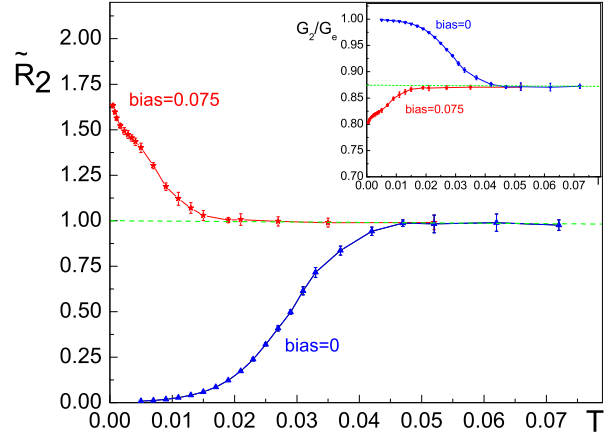}
\end{center}
\caption{(Color online) $\tilde{R}_2$ linear ($F=0.001$) response of non-slanted (lower curve) and slanted (upper curve) dislocations. Parameters are $L_x=80, U_C=2, K=1,V_d=1$; bias is determined as linear density of extra kinks. Inset shows corresponding $G_2 $-modulus determined from Eq.(\ref{G122}) with $\gamma=0.15$. Slanted dislocations induce low-$T$ shear modulus softening shown by the lower curve in the inset. Green gashed line indicates response of free Granato-L\"ucke string \cite{Granato}.}
\label{bias}
\end{figure}  
\begin{figure}
\begin{center}
\includegraphics[%
  width=0.65\linewidth,
  keepaspectratio]{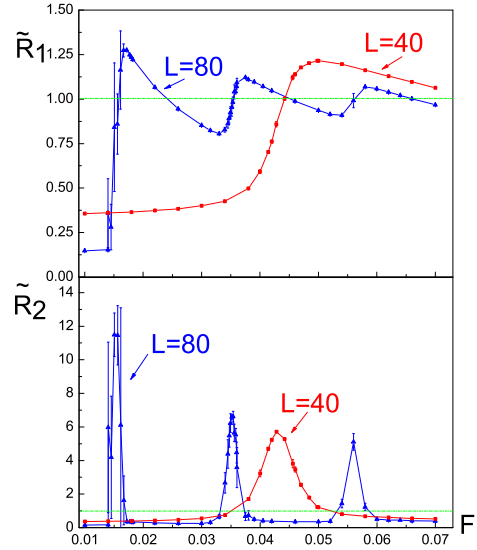}
\end{center}
\caption{(Color online) Full $\tilde{R}_1$ and differential $\tilde{R}_2$ responses of gliding dislocations of different lengths on applied external stress $\sigma$ ($F=b\sigma$) at temperature $T=0.02$.Green dashed lines indicate responses of free strings (in the absence of Peierls potential). Error bars are shown for every point. Lines are guides to eyes. Note that the threshold $F_c$ for $F$ (sharp increase of $\tilde{R}_1$ and peak in $\tilde{R}_2$) is about two times smaller for $L_x=80$ than that for $L_x=40$. Next peaks positions are approximately periodic with the period $\sim F_c$.}
\label{compare}
\end{figure}
\begin{figure}
\begin{center}
\includegraphics[%
  width=0.65\linewidth,
  keepaspectratio]{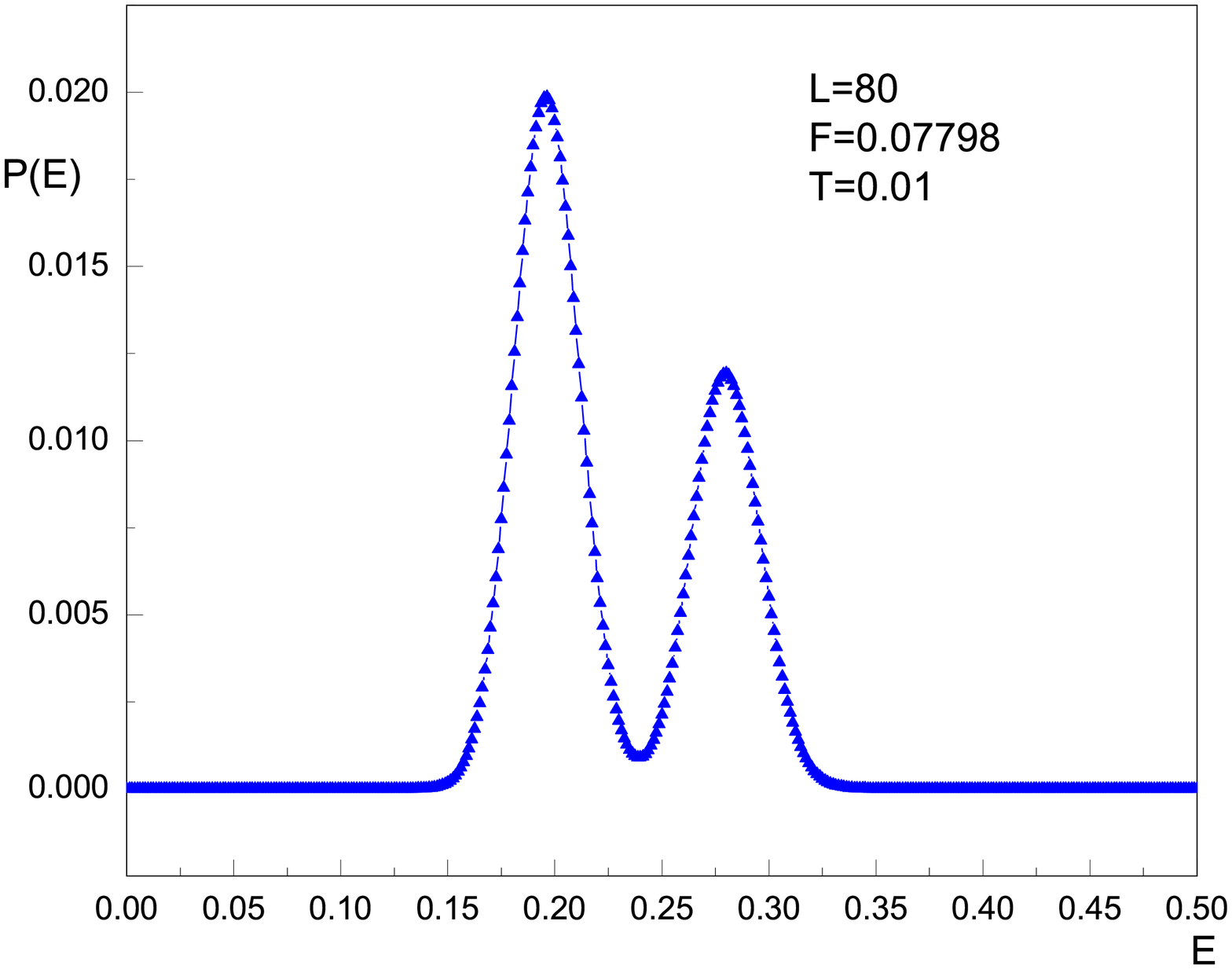}
\end{center}
\caption{(Color online) Energy histogram for gliding dislocation undergoing low-T stress induced roughening. Bimodality of $P(E)$ is a clear signature of jump-like transition and hysteretic behavior. }
\label{bimodal}
\end{figure}

As proven in Ref.\cite{EPL}, at $T=0$ gliding non-slanted dislocation must be pinned by Peierls potential. This implies $\tilde{R}_{1,2} ~ n_d <<1$ so that $G_{1,2}=G_e$. As $T$ increases thermal roughening takes place due to creation of kink-antikink pairs. This leads to increase of $\tilde{R}_{1,2}$ so that eventually $\tilde{R}_{1,2} \to 1$ and the moduli $G_{1,2}$ attain their high-T value $G_{\infty}$. Ref.\cite{EPL}has focused on describing this effect in the limit of no bias ($y_{max}=0$) and vanishing external stress ($ F\to 0$).

Here we address the question how  finite stress and bias affect the roughening process. 
In general, roughening could be expected to be achieved by increasing temperature, bias and applied stress. All three factors tend to introduce kinks which may be expected to behave as free quantum particles which liberate dislocation from Peierls potential. However, as we will discuss below, these three factors lead to different scenarios.

\subsection{Finite stress}
 Thermal roughening at small stresses is a relatively smooth crossover \cite{EPL}. It is demonstrated by lower curve in Fig.\ref{bias}. In contrast, inducing roughening by increasing stress is characterized by non-monotonous, jump-like features. Fig.\ref{compare} demonstrates such behavior for $R_1$ and $R_2$ at constant temperature $T=0.02$ (in dimensionless units). It is important that the threshold stress (the parameter $F$), inducing such features,  depends essentially on the dislocation length. The non-linearity is associated with work $FL_x$ done by stress to create a single kink-antikink pair carrying energy $\Delta$ \cite{Hirth}. Further increase of $F$ creates more kinks in a quasi-periodic fashion  with the overall tendency to approach free string response (green dashed line). This behavior is clearly   seen for $L_x=80$ case in Fig.\ref{compare}. 

\begin{figure}
\begin{center}
\includegraphics[%
  width=0.65\linewidth,
  keepaspectratio]{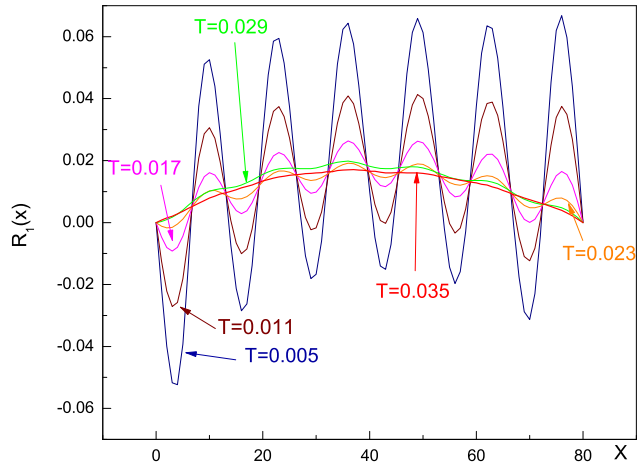}
\end{center}
\caption{(Color online) Spatial correlator $R_1(x)$ (related to $\tilde{R}_1$ as $\tilde{R}_1=\sum_x R_1(x)$) of the biased string with the same parameters as in Fig.\ref{bias} at different $T$ . Spatial modulations reflect Wigner-type crystallization of the extra kinks. At high $T$, the spatial profile becomes that of the Granato-L\"ucke string pinned at both ends \cite{Granato}.}
\label{C1}
\end{figure}  

We have found that force induced roughening proceeds similarly to I-order-transition at low $T$. Fig.\ref{bimodal} demonstrates bimodal distribution of energy for dislocation undergoing stress induced roughening at $T=0.01$. Clearly, such bimodality implies hysteresis in mechanical properties. The bimodality vanishes above certain temperature. For chosen parameters $U_C=2, u_p=0.1, K=1, L_x=80$ this occurs above $T\approx 0.015$.
More systematic studies how the onset temperature of hysteretic behavior depends on dislocation parameters will be performed elsewhere. 

%Differences in reponses $R_1$ and $R_2$ can be seen in $T$-dependencies at finite foces. Fig.\ref{F0.03} demsontrtes it for dislocation of length $L_x=40$.For comparison, Fig.\ref{F0.01} demonstrates thermal roughening in the presence of small force (in the regime studied in Ref.\cite{EPL})

\subsection{Finite Bias and shear modulus softening at low $T$}
Finite bias (slanting) is relevant to the situation of strong pinning of a dislocation at its ends in different vallyes of Peierls potential. Otherwise, dislocation will lower its energy by settling into a single Peierls valley.
Strong end pinning with slanting imposes geometrical constraint and, therefore, introduces kinks even at $T=0$. Such kinks change the responses dramatically: dislocation remains rough at $T$ much lower than that in the absence of bias. Fig.\ref{bias} demonstrates this effect in $\tilde{R}_2$ response of non-slanted dislocation (lower blue curve) and of the slanted one (red upper curve). The first is smooth at low $T$ and becomes thermally rough at elevated $T$. In contrast, the second, which has six geometrical kinks (given $L_x=80$, the bias is $6/80=0.075$), exhibits just the opposite behavior -- softening at low $T$. Accordingly, shear modulus $G_2$, Eq.(\ref{G122}, (see the inset in Fig.\ref{bias})  hardens at low $T$ for non-slanted dislocations and softens for slanted ones. Very recently such behavior has been reported in Ref.\cite{Balibar3} for impurities free crystals of \he4: shear modulus of \he4 crystals grown at $20$mK from superfluid has exhibited significantly smaller values than that at high $T$. Such softening, however, vanishes after the crystal is warmed up and cooled down again. According to Ref.\cite{Balibar3}, this happens due to \hee3 reentering solid \he4 and binding dislocations. 

We attribute such drastically different behavior of slanted dislocations in impurities free crystal to quantum nature of the extra kinks. Loosely speaking they form "superfluid" which enhances dislocation mobility at low $T$. The long-range potential (\ref{C}) does not change significantly this picture in spite of inducing Wigner-type crystal quasi-order among extra kinks. Such order is seen in the spatial oscillations of the correlation function $R_1(x)=\langle y(x) \rangle/(R_0F)$, Fig.\ref{C1}. In 1d achieving true crystal order is impossible, and, therefore, kinks remain "superfluid" and "solid" at the same time -- some sort of "supersolid" of the extra kinks. 

However, dislocation network resides in 3d sample, and actual Wigner crystallization of kinks should be possible. This implies that shear modulus may, first, demonstrate softening as $T$ lowers and, then, at $T<T^*$, -- hardening and eventually the recovery to its ideal value $G_e$. If, however, $n_d$ is very low, such recovery may never be observed. In order to estimate $T^*$ we compare it with typical "Coulomb" interaction $\propto U_C/r$ between kinks at their average separation $r\propto 1/\sqrt{N_d}$, provided their typical zero-point kinetic energy is smaller than potential. This gives $T^*\sim T_1 \sqrt{n_d}$, where $T_1$ stands for a typical energy scale $10$K in \he4. For dislocation densities $N_d\sim 10^{10}-10^{12}$m$^{-2}$ (the dimensionless density $n_d \sim 10^{-9}-10^{-7}$) we find $T^*\sim 1-10$mK. This should be compared with the temperature of thermal roughening $T_r$ determined by typical energy of  kink-antikink pair creation, which is a single dislocation effect and therefore is independent of $n_d$. As discussed in Ref.\cite{EPL},  $T_r\sim 100-200$mK.

\section{Superclimb}
In the experiment \cite{Ray} pressure  applied to  superfluid \he4 inside the vycor "electrode" was transferred to the solid \he4 through relatively small contact area.  Yet, this caused large {\it uniform} (which is inconsistent with plastic flow) accumulation of matter inside the solid. In this sense the effect of pressure can be viewed as a mean to change chemical potential $\mu$.  
This phenomenon was called \textit{giant isochoric} compressibility \cite{sclimb} -- macroscopic accumulation of \he4 atoms in response on changing $\mu$.
It has been proposed that the actual matter accumulation occurs due to building half-planes forming superfluid edge dislocations.
\begin{figure}
\begin{center}
\includegraphics[%
  width=0.65\linewidth,
  keepaspectratio]{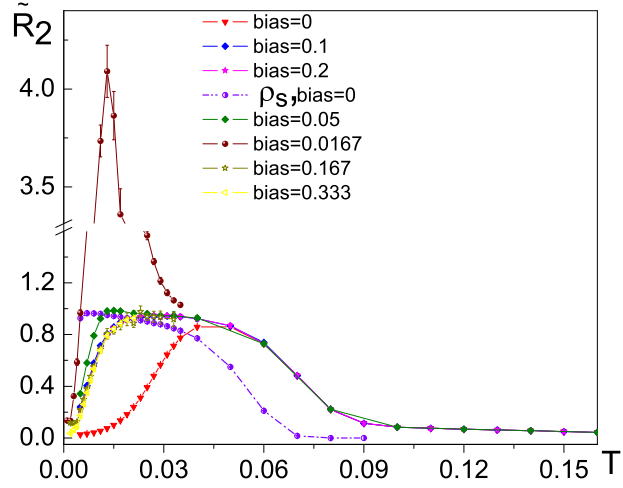}
\end{center}
\caption{(Color online) Temperature dependencies of $\tilde{R}_2$ (with $\chi_2 \propto \tilde{R}_2$, Eq.(\ref{Chi})) for superclimbing dislocation for different biases $y_{max}/L_x$; $F=0.001$, $L_x=60$, $\rho_0(0)=1,\, T_0=0.2$, $K=1$, $U_C=1$.  Dot-dashed line represents renormalized superfluid stiffness $\rho_s(T)$, with the bare one taken in the form $\rho_0(T)=\rho_0(0)(1-T/T_0)^\nu,\, \nu=0.67$.
Anomalously large response is found for one extra jog, bias=$1/60\approx 0.0167$.}
\label{R2}
\end{figure}  
\begin{figure}
\begin{center}
\includegraphics[%
  width=0.65\linewidth,
  keepaspectratio]{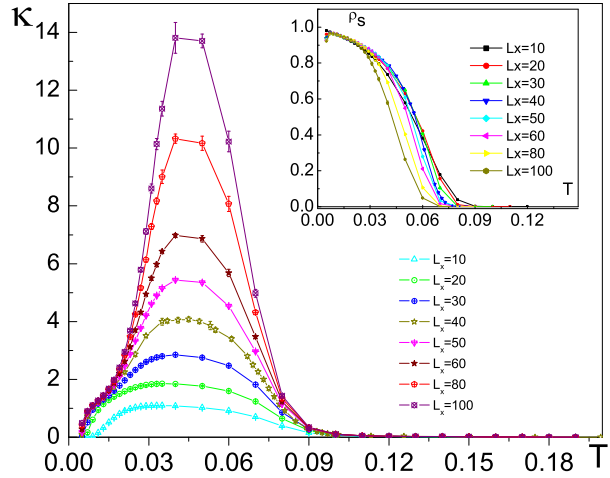}
\end{center}
\caption{(Color online) Temperature and size dependencies of $\kappa(T,L_x)$ for superclimbing dislocation at zero bias. Size dependence of the peak ($T=0.05$) can well be fit by the power law  $\kappa(T=0.05,L_x) \propto L_x^w,\,\, w\approx 1.32$ for $U_C=1$.Inset shows $\rho_s(T,L_x)$. It becomes suppressed drastically at about $1/2$ of $T_0=0.2$ -- mean field temperature where microscopic superfluidity vanishes. Size dependence of $\rho_s$ is insignificant if compared with $\kappa$.}
\label{kappa}
\end{figure} 
 
We consider two isochoric compressibilities 
\begin{eqnarray}
\chi_1= \frac{\Delta N}{N\Delta \mu}, \quad \chi_2 = \frac{dN}{Nd\mu}
\label{Chi12}
\end{eqnarray}
 where $\Delta N$ is the amount of \he4 atoms accumulated in the sample upon applying chemical potential difference $\Delta \mu$, with $N$ being total number of atoms in the sample. Variation of $\mu$ can be identified with deviation 
of local pressure  $p$ from the equilibrium value $p_0$ as $\Delta \mu = (p-p_0)\Omega$ where $\Omega $ stands for volume occupied by one atom. Force $F$ in Eq.(\ref{Hdis})
is given by $F=(p-p_0)b= \Delta \mu b/\Omega$. Thus, $d.../d\mu = (b/\Omega) d.../dF$. $\Delta N$ can be related to average (dimensionless) density of climbing dislocations $n_d$. Indeed, if the dislocation forest is characterized by typical length $L_x$ of smallest free segment, its displacement $y$ changes number of atoms by $ \approx 3 y L b$  per each block. Thus, the fraction of atoms changed is $ \approx 3 y b L_x/L_x^3 \approx n_d y/b$. This gives $\chi_{1,2} \propto n_d R_{1,2}$. 
For free segments of dislocations the responses $R_{1,2}$ become $R_0 \propto L_x^2$ which implies compressibilities comparable to liquid \he4 $\chi_0$ (since $n_d \propto 1/L_x^2$). Thus, introducing
relative compressibilities $\tilde{\chi}_{1,2}=\chi_{1,2}/\chi_0$ and $\tilde{R}_{1,2}=R_{1,2}/R_0$, one  finds 
in the dimensionless units
\begin{eqnarray}
\tilde{\chi}_{1,2} = \gamma \tilde{R}_{1,2},
\label{Chi}
\end{eqnarray}
that is, $\chi_{1,2} \propto \tilde{R}_{1,2}$,
where we ignore compressibility due to direct application of pressure to the area of contact between "electrodes" \cite{Ray} and solid \he4 sample.
The parameter $\gamma \sim 0.1-1$ is the geometrical factor depending on the structure of the dislocation forest. 

The responses $\tilde{R}_{1,2}$ have been determined from Monte Carlo simulations of the full action (\ref{H}-\ref{or}). In this paper we consider limit of small $F$ only (where two responses coincide). Non-linearities in the superclimb effect will be addressed elsewhere.
As discussed in Ref.\cite{sclimb}, Peierls potential with respect to jog formation is always relevant at low $T$, and, therefore, the compressibility will be suppressed.
This is demonstrated in Fig.\ref{R2} where the response $\tilde{R}_2$ becomes significantly suppressed at low $T$ for, practically, any bias. 
As $T$ rises, jogs are created thermally, provided the core superflow can still exist, making Peierls potential less and less relevant. Accordingly, $\tilde{R}_{1,2}$ should approach unity (that is, the free string limit $R_0\propto L_x^2$). 
At higher $T$ superfluidity ceases to exist and so does the giant compressibility effect. 

As discussed in Ref.\cite{Shevchenko}, dislocation network superfluidity is characterized by two temperature scales -- microscopic $T_0 \sim 1$K (comparable with the bulk $\lambda$-temperature) and macroscopic $T_c\propto T_0 a/L_x << T_0$. In our model, we choose the bare superfluid stiffness  in Eq.(\ref{HSF}) to vanish according 
to the 3D XY-model law $\rho_0(T)=\rho_0(0)(1-T/T_0)^\nu, \,\, \nu \approx 0.67$ with the choice $T_0=0.2$ (in dimensionless units). As can be seen in Fig.\ref{R2}, high-$T$ behavior of $R_2$ reflects suppression of the renormalized superfluid stiffness $\rho_s(T)$ which happens at about 1/2 of $T_0$. Strictly speaking, the 3d form of the bare stiffness is not valid very close to $T_0$ and one should rather use the mean field form. However, since major suppression occurs at much lower $T$, the actual form of $\rho_0(T)$ is not important.

Another prediction for the superclimb effect is that the core superfluidity becomes of non-Luttinger type with the free-particle-type dispersion \cite{sclimb} for $U_C=0$.
In terms of the superfluid compressibility $\kappa$ this implies strong dependence $\kappa \propto L_x^2$ on the dislocation length in the temperature range where superclimb is observed. We have measured superfluid compressibility through imaginary time windings fluctuations of the J-currents $m_t$. Fig.\ref{kappa} presents temperature dependence of $\kappa$ for dislocation lengths $L_x=10-100$ at zero bias. At low $T$ all curves collapse, practically, into a single one. This indicates the region of the standard Luttinger liquid behavior. As $T$ rises, curves for different $L_x$ diverge, with their maxima obeying asymptotic behavior $\kappa_{max} \propto L_x^w, \,\, w\approx 1.32$ for $ U_C=1$. In general, the exponent $w$ depends on long-range interaction strength $U_C$, approaching $w=2$ for $U_C=0$. At higher $T$ the curves collapse again and eventually $\kappa$ vanishes.

We have found that bias with respect to climb does not render Peierls potential irrelevant at $T=0$. It simply lowers the temperature when $\chi_{1,2}$ become suppressed. As can be seen from Fig.\ref{R2}, $\chi_2$ at finite biases starts diminishing with lowering $T$ at approximately 3 times smaller temperature than that for zero bias. Qualitatively, this scenario has been found to be independent of the "Coulomb" interaction (\ref{C}) between jogs.

\section{Conclusions}
 Low-$T$ dislocation glide is characterized by hysteretic behavior and ratchet-type shape of $R_1$-response as a function of applied stress, with the $R_2$-response exhibiting peaks at the steepest parts of the "ratchet". The threshold stress is determined by Peierls gap and it vanishes inversely proportional to approximately dislocation length. Thus, these type of measurements may reveal temperature dependence of Peierls gap as well as give an estimate of a typical length of dislocation segments between strongly pinned crosslinking points \cite{note}. 

Geometrical bias (due to ends of dislocation residing in different Peierls potential valleys) introduces kinks even at $T=0$ and produces  dramatic effect on gliding dislocation: while at zero bias dislocation is always quantum smooth and shows thermal roughening above some temperature $T_r$ \cite{EPL}, single biased dislocation
remains rough at $T<<T_r$. Furthemore, effective rigidity of such dislocation shows softening with respect to its value at high $T$. If such biased dislocations dominate
the network, shear modulus should exhibit softening at low $T$ \cite{Balibar3} -- the effect opposite to the observation \cite{Beamish}. 
Long-range forces between kinks lead to Wigner-type quasi-crystalization of kinks. This effect, however, does not affect significantly quantum roughening of biased dislocation due to its 1d nature. At finite density of dislocations, true Wigner crystalization of kinks should take place below temperatures determined by dislocation density when the system becomes essentially 3d.
Such softening can be suppressed by \hee3 impurities localizing the extra kinks.
Thus, it is important to study shear moduli in impurities free solid \he4 crystals as suggested in Ref.\cite{Balibar2} in order to reveal the nature of dislocation crosslinking. Observing low-$T$ moduli softening insensitive to thermal cycling would imply significant presence of slanted dislocations stabilized by strong pinning at the crosslinking. Conversely, observing hardening induced by annealing would mean that the dislocation forest is able to reach more stable configuration without geometrically imposed zero-point kinks.
As recent report \cite{Balibar3} indicates, despite restoring the low-$T$ hardening after thermal cycling, there is very slow relaxation toward the low-$T$ soft state due to escape of \hee3  back into the liquid which is in a contact with the solid. This favors the first scenario -- stable dislocation network dominated by slanted dislocations.

Isochoric compressibility due to superclimbing dislocations has been evaluated and the prediction of its low-$T$ suppression \cite{sclimb} has been confirmed. In contrast to glide, such suppression is not qualitatively affected by geometrical slanting.  We attribute this to the nature of jog superclimbing dynamics -- strong control of jog displacement by supercurrents. More specifically, in order to displace $N_j$ jogs by some distance $S$ along the dislocation, $SN_j$ atoms must be transferred through  its ends. Two factors inhibit such transfer -- interaction between (extra) jogs and low-$T$ suppression of compressibility  due to Peierls barrier far from jogs (where bias is irrelevant). Single jog is not affected by the interaction and, thus, can move freely to induce large fluctuations of the total number of atoms, that is, -- large compressibility. At low $T$ the second factor suppresses this anomalous single-jog behavior. An interesting subject for future studies is non-linearities of the superclimb with respect to applied pressure.

\begin{acknowledgements}
We would like to thank Sebastien Balibar, John Beamish, David Schmeltzer and Boris Svistunov for stimulating discussions. We acknowledge support from the National Science Foundation, NSF grant PHY-0653135, and CUNY PSC grant as well as CSI-CUNY  supercomputing facility.
\end{acknowledgements}

%\pagebreak


\begin{thebibliography}{99}

\bibitem{Andreev69} A. F. Andreev and I. M. Lifshitz, Sov. Phys. JETP {\bf 29}, 1107 (1969); 
D. J. Thouless %,The flow of a dense superfluid, Ann. Phys. {\bf 52}, 403-427(1969) 
Ann. Phys. {\bf 52}, 403(1969)
G. V. Chester, Phys. Rev. A {\bf 2}, 256 (1970).

\bibitem{KC} E. Kim and M.H.W. Chan, Nature {\bf 427}, 225
(2004).
\bibitem{KC2}
 E. Kim and M.H.W. Chan,Science {\bf 305}, 1941 (2004).

\bibitem{noSFS}
M. Boninsegni, et. al., %N. Prokof'ev and B. Svistunov, 
Phys. Rev. Lett. {\bf 96}, 105301 (2006);
%\bibitem{Boninsegni06a} Boninsegni, M. , Prokof'ev, N. \& Svistunov B. Superglass phase of ${}^4$He. {\it Phys. Rev. Lett.} {\bf 96}, 105301 (2006).
B. K. Clark and D. M. Ceperley, Phys. Rev. Lett. {\bf 96}, 105302 (2006);
%\bibitem{Clark06} Clark, B. K. \& Ceperley, D. M. Off-diagonal long-range order in solid ${}^4$He. {\it Phys. Rev. Lett.} {\bf 96}, 105302 (2006).
M. Boninsegni, et. al., % A. B. Kuklov, L. Pollet, N. V. Prokof'ev, B. V. Svistunov, and M. Troyer, 
Phys. Rev. Lett. {\bf 97}, 080401 (2006); 
%\bibitem{Boninsegni07} Boninsegni, M.,  Kuklov, A. B., Pollet, L., Prokof'ev, N. V., Svistunov, B. V. \& Troyer, M. Luttinger liquid in the core of a screw dislocation in Helium-4. {\it Phys. Rev. Lett.} {\bf 99}, 035301 (2007)..
%\bibitem{Boninsegni06} Boninsegni, M.,  Kuklov, A. B., Pollet, L., Prokof'ev, N. V., Svistunov, B. V. \& Troyer, M. Fate of vacancy-induced supersolidity in ${}^4$He. {\it Phys. Rev. Lett. } {\bf 97}, 080401 (2006).

\bibitem{Balibar} S. Balibar and F. Caupin, J. Phys. Cond. Mat. {\bf 20}, 173201 (2008).

\bibitem{screw}
M. Boninsegni, A. B. Kuklov, L. Pollet, N.V. Prokof'ev, B.V. Svistunov, \& M. Troyer, 
%M. Luttinger liquid in the core of a screw dislocation in Helium-4. 
{\it Phys. Rev. Lett.} {\bf 99}, 035301 (2007);
\bibitem{stress}
L. Pollet, M. Boninsegni, A.B. Kuklov, N.V. Prokof'ev,B.V. Svistunov, and M. Troyer, Phys. Rev. Lett. {\bf 101}, 097202 (2008);
 %Local Stress and Superfluid Properties of Solid 4He  
Phys. Rev. Lett. {\bf 101}, 269901 (2008).
\bibitem{sclimb} S. G. S\"oyler, A. B. Kuklov, L. Pollet, N.V. Prokof'ev,and B.V. Svistunov,
Phys.Rev.Lett. {\bf 103}, 175301 (2009).

\bibitem{Friedel} Friedel J., {\it Dislocations}, Pergamon Press, 1964.
\bibitem{Beamish}
J. Day and J. Beamish, Nature {\bf 450}, 853 (2007);
 J. Day,O. Syshchenko, and J. Beamish,Phys. Rev. {\bf B 79}, 214524(2009). 
%Intrinsic and dislocation-induced elastic behavior of solid helium
\bibitem{Balibar3} S. Balibar,X. Rojas, H. Maris, 2010 March Meeting, talk on March 18, 2010, Abstract: V23.00012;
X. Rojas, A. Haziot, V. Bapst, S. Balibar, H.J. Maris, preprint (unpublished).

\bibitem{Ray} M. W. Ray and R. B. Hallock, %Observation of Unusual Mass Transport in Solid hcp 4He
Phys. Rev. Lett. {\bf 100}, 235301 (2008); %Observation of mass transport through solid  4He
Phys. Rev. B {\bf 79}, 224302 (2009).  

\bibitem{EPL} 
D. Aleinikava, E. Dedits, A. B. Kuklov and D. Schmeltzer, %Dislocation roughening in quantum crystals
Europhys. Lett. , {\bf 89}  46002 (2010); arXiv:0812.0983



\bibitem{Wallin} M. Wallin,E.S. S\"orensen,S. M. Girvin, A. P. Young, Phys. Rev. {\bf B 49}, 12115 % - 12139 
(1994)

\bibitem{WA} N.V. Prokof'ev and B.V. Svistunov, Phys. Rev. Lett. {\bf 87}, 160601 (2001). 

\bibitem{Hirth1} Hirth J. P. and Lothe J., Theory of Dislocations
(McGraw-Hill), 1968; Kosevich A. M., The Crystal Lattice: Phonons, Solitons,
Dislocations, Superlattices (Wiley) 2005.

\bibitem{Granato} A. Granato and K. L\"ucke, J. Appl. Phys. {\bf 27}, 583 (1956); ibid. 789(1956).

\bibitem{Hirth} J. Lothe and J. P. Hirth, Phys. Rev. {\bf 115}, 543 (1959) ; Petukhov B. V. and Pokrovskii V. L., JETP, {\bf 36}
(1973) 336.  	
%Dislocation Dynamics at Low Temperatures 

\bibitem{Shevchenko} S. I. Shevchenko, Sov. J. Low Temp. Phys. {\bf 13}, 61 (1987).
%\bibitem{Schevchenko}  Shevchenko, S. I. One-dimensional superfluidity in Bose crystals. {\it Sov. J. Low Temp. Phys. } {\bf 13}, 61 (1987).

\bibitem{Balibar2} Rojas X., Pantalei C., Maris H. J. and Balibar S.,
J. Low Temp. Phys., {\bf 158} , 478 (2010).

\bibitem{note} Extremely small stresses for the onset of non-linearity observed in Ref.\cite{Beamish} may hint on the possibility that there is actually no 
strong pinning due to crosslinking, and that the  length of dislocation determining the threshold is comparable to sample size. 

\end{thebibliography}
\end{document}